\documentclass[journal,twocolumn]{IEEEtran}
\usepackage{epsfig,makeidx,color}
\usepackage{amsmath,amssymb,bbm}
\usepackage{cite,graphicx,lipsum}
\usepackage{enumerate}
\usepackage[switch,pagewise]{lineno}
\usepackage{hyperref}
\hypersetup{
        colorlinks = true,
        citecolor=blue,
}
\newcommand{\indicator}{\mathbbm{1}}

\def\cQ{{\cal Q}}

\def\rH{{\rm H}}
\def\rT{{\rm T}}

\def\uE{{\mathbb E}}

\DeclareMathOperator*{\argmin}{\arg\!\min}

\newtheorem{mylemma}{\bf Lemma} 

\def\deft{ \buildrel \triangle \over = }

\def\be{ \begin{equation} }
\def\ee{ \end{equation} }
\def\bea{ \begin{eqnarray} }
\def\eea{ \end{eqnarray} }

\def\bx{{\bf x}}

\def\ba{{\bf a}}
\def\br{{\bf r}}

\def\bn{{\bf n}}

\def\bI{{\bf I}}

\def\bR{{\bf R}}

\def\b0{{\bf 0}}

\def\cA{{\cal A}}
\def\cC{{\cal C}}

\def\cQ{{\cal Q}}

\def\cN{{\cal N}}

\ifCLASSOPTIONonecolumn
  \newcommand{\figwidth}{0.50\columnwidth}
\else
  \newcommand{\figwidth}{0.90\columnwidth}
\fi

\begin{document}

\title{Opportunistic NOMA for Uplink Short-Message
Delivery with a Delay Constraint}

\author{Jinho Choi\\
\thanks{The author is with
the School of Information Technology,
Deakin University, Geelong, VIC 3220, Australia
(e-mail: jinho.choi@deakin.edu.au).
This research was supported
by the Australian Government through the Australian Research
Council's Discovery Projects funding scheme (DP200100391).}}


\maketitle
\begin{abstract}
In this paper, we study the application of
opportunistic non-orthogonal multiple
access (NOMA) mode to short-message transmissions
with user's power control under a finite power budget.
It is shown that 
opportunistic NOMA mode, which can
transmit multiple packets per slot, can dramatically
lower the session error probability
when $W$ packets are to be transmitted within 
a session consisting of $W_{\rm S}$ slots,
where $W_{\rm S} \ge W$ and the slot length is equivalent
to the packet length, compared to orthogonal multiple access (OMA)
where at most one packet can be transmitted in each slot.
From this, 
opportunistic NOMA mode can be seen as an
attractive approach for uplink transmissions.
We derive an upper-bound on the session error probability
as a closed-form expression and also obtain a closed-form for the NOMA
factor that shows the minimum 
possible ratio of the session error probability of opportunistic NOMA to
that of OMA. Simulation results also confirm that 
opportunistic NOMA mode has a much lower session error probability
than OMA.
\end{abstract}


\ifCLASSOPTIONonecolumn
\baselineskip 28pt
\fi

\section{Introduction}

Non-orthogonal multiple access (NOMA)
has been extensively studied 
for cellular systems \cite{Dai15} \cite{Ding_CM}
\cite{Choi17_ISWCS} \cite{Choi17_JCN} \cite{Dai18},
because it can provide a higher spectral 
efficiency than orthogonal multiple access (OMA).
In particular, 
in order to support multiple users with the same radio resource block
for downlink with beamforming in cellular systems,
power-domain NOMA with
successive interference cancellation (SIC) is considered
\cite{Saito13} \cite{Kim13}.
It is also possible to employ 
the notion of NOMA for reliable
transmissions in ad-hoc networks as in \cite{Choi08G}.
In \cite{Choi17}, a power control
policy for NOMA is studied with a delay constraint.

The Internet of Things (IoT) 
has attracted enormous attention in recent years.
There are a large number of IoT applications including
smart factory and smart cities \cite{Shrouf14} \cite{Dixit15} \cite{Lom16}.
In most cases, IoT devices and sensors
are connected to the Internet and expected to upload
their data or measurements,
which are mainly short messages \cite{Shafiq12} \cite{Sehati18}.
For real-time applications, it is expected
that short messages can be delivered within a certain delay limit
\cite{Naik17} \cite{Kim18}.

In this paper, we consider uplink
transmissions where a finite number of 
packets are to be delivered within a certain number of slots
with a high probability,
which is referred to as short-message delivery 
with a delay constraint (SMDDC).
To this end, 
the channel state information (CSI) of users is required at 
a base station (BS) to perform full resource allocation.
The BS can estimate the 
instantaneous CSI of time-varying fading channels
for a packet from a user 
when a packet includes a (short) pilot
sequence to allow the channel estimation \cite{TseBook05}
in time division duplexing (TDD) mode thanks to the channel reciprocity.
In downlink transmissions, 
the BS with known CSI can perform resource allocation
and power control for dowlink packets 
in order to guarantee target performances.
However, in uplink transmissions,
it is difficult for the BS to 
perform resource allocation to guarantee target performances,
because the CSI of users at the BS might be outdated
over fast fading channels when users transmit packets.
Consequently, throughout the paper, we assume that
the BS only allocates the radio resource blocks to 
users, while users perform their power control with known their CSI 
and a limited power budget for NOMA.
Note that since short message is considered in this paper,
the approach differs from that in \cite{Choi17},
where a continuous data stream is assumed with a queue,
and it might be more suitable for
IoT applications where short messages from IoT devices or sensors
are to be delivered within a certain time limit.

For SMDDC, we assume that each user has a finite number of
packets to be transmitted within a certain time and
one packet can be transmitted within a slot in this paper.
Thus, if there is no decoding failure at the BS, 
a user needs to have $W$ slots to deliver $W$ packets
(usually, $W$ may not be too large in mission-critical
applications, e.g., remote surgery, with SMDDC).
However, due to deep fading, with a limited power budget,
a user may not be able to successfully transmit some packets.
Therefore, in order to complete the delivery of $W$ packets,
in general, we need more than $W$ slots, say $W_{\rm S}$ slots,
where $W_{\rm S} \ge W$.
Thus, for SMDDC, it is expected that 
$W_{\rm S}$ is sufficiently small 
with a high probability that all $W$ packets can be successfully
transmitted within $W_{\rm S}$ slots.

In this paper, we show that for OMA, $W_{\rm S}$ cannot be
close to $W$ (under Rayleigh fading) 
with a high probability
of successful transmissions of $W$ packets. 
However, using opportunistic NOMA mode,
which is proposed
in this paper 
to transmit more than one packet per slot
using others' channels based on power-domain NOMA,
$W_{\rm S}$ can be close $W$ 
with a high probability
of successful transmissions of $W$ packets. 
For example, under Rayleigh fading, 
the probability of successful transmissions of $W = 50$ packets 
becomes about $1 - 10^{-4}$ with $W_{\rm S} = 60$ slots 
if a proposed NOMA scheme is used. On the other hand,
the probability of successful transmissions of $W = 50$ packets 
becomes more than 0.5 if OMA is used.

In summary, the main contributions of the paper
are two-fold: 
\emph{i}) opportunistic NOMA
schemes are proposed for SMDDC;
\emph{ii}) a closed-form expression for
an upper-bound on the session error probability
(which will be defined later)
is derived to see the impact of NOMA on 
the session error probability under independent Rayleigh fading
(as well as a closed-form for the NOMA factor).

The rest of the paper is organized as follows.
In Section~\ref{S:SM}, we present the system model
for SMDDC based on
the notion of (power-domain) NOMA.
The power allocation is studied in Section~\ref{S:NOMA}
with a limited power budget.
The probabilities of multi-packet transmissions
by different NOMA schemes
are considered and their closed-form expressions
are derived under independent Rayleigh fading in
Section~\ref{S:PMP}.
With a scenario for SMDDC,
the session error probability is defined and its
upper-bound is derived in Section~\ref{S:OA}.
Simulation results are presented in Section~\ref{S:Sim} and
the paper is concluded with some remarks in Section~\ref{S:Conc}.

\subsubsection*{Notation}
Matrices and
vectors are denoted by upper- and lower-case
boldface letters, respectively.
The superscripts $\rT$ and $\rH$
denote the transpose and complex conjugate, respectively.
The Kronecker delta is denoted by $\delta_{l,l^\prime}$,
which is 1 if $l = l^\prime$ and 0 otherwise.
$\uE[\cdot]$
and ${\rm Var}(\cdot)$
denote the statistical expectation and variance, respectively.
$\cC\cN(\ba, \bR)$
represents the distribution of
circularly symmetric complex Gaussian (CSCG)
random vectors with mean vector $\ba$ and
covariance matrix $\bR$.
The Q-function is given by
$\cQ(x) = \int_x^\infty \frac{1}{\sqrt{2 \pi} } e^{- \frac{t^2}{2} } dt$.

\section{System Models}	\label{S:SM}


In this section, we assume that there are 
$M$ orthogonal radio resource blocks or (multiple access) channels
(in the frequency or code domain) for uplink transmissions
with $K$ users assigned to $M$ channels.
We first present 
OMA where each user is allocated to a dedicated channel
with $M = K$. 
Then, we present two different approaches for opportunistic NOMA.
For easy comparisons, we also assume that $M = K$ in NOMA
and show that a user can opportunistically use 
other channels to transmit more than one packet 
with different power levels (for
successful SIC at a BS).

Throughout the paper, let $h_{m,k}(t)$ denote
the channel coefficient
from user $k$ through the $m$th radio resource block 
at time slot $t$. In addition, 
we assume block fading channels \cite{TseBook05},
where $h_{m,k} (t)$ remains unchanged over the duration of a slot and
randomly varies from a time slot to another.

\subsection{OMA System}

In OMA, we have
$K = M$ so that each user can have one 
(orthogonal) radio resource block for uplink transmissions.
Furthermore, assume that
the $m$th radio resource block
is assigned to user $m$, i.e., $k = m$. 
Then, letting $\br_m (t)$ represent the received signal at the BS 
through the $m$th radio resource block or channel
at time slot $t$, we have
$$
\br_m (t) = h_{m,m} (t) \ba_m (t) + \bn_m (t),
\ m =1, \ldots, M,
$$
where 
$\ba_m (t)$ is the packet transmitted by user $m$,
and
$\bn_m (t) \sim \cC \cN(0, N_0 \bI)$ is the background noise 
in the $m$th channel at the BS.
If a user's packets are not successfully transmitted
(e.g., due to decoding errors at the BS under deep fading),
there should be re-transmissions through the same radio
resource block using 
HARQ protocols for reliable transmissions \cite{LinBook}, 
which results in delay.

If short packets are considered, the overhead of feedback signals
for HARQ to each user might be high. 
To avoid a high feedback overhead, we can consider
the power control at users with \emph{known} channel coefficients.
That is, each user can decide the transmit power of 
packets to meet the required signal-to-noise ratio (SNR) or 
signal-to-interference-plus-noise ratio (SINR) for successful decoding.
To this end, throughout the paper, 
we assume time-division duplexing (TDD) mode. 
The BS transmits a beacon signal prior to 
each slot so that all the users can estimate their channel coefficients
to the BS thanks to the channel reciprocity and perform
power control.
In this case, a user cannot transmit a packet when
the channel gain is not sufficiently high (to meet 
the required SINR with a limited power budget), which leads to delay.

In order to avoid a long delay,
the notion of NOMA can be used in an 
opportunistic manner to transmit more than one packet per slot.
There are two different systems to apply
opportunistic NOMA mode to uplink, which are discussed below.

\subsection{Symmetric NOMA System}

Since there are $M$ channels,
a user can transmit $M$ packets simultaneously if necessary.
Thus,
for example, if user $k$ has $M-1$ additional packets to 
transmit, the received signals at the BS from user $k$
are given by
$$
\bx_{m(k,i)} (t) = h_{m(k,i),k} (t) \ba_k (t;i),
\ i \in \{1, \ldots, M\},
$$
where $m(k,i) \in \{1, \ldots, M\}$ 
represents the index of channel (or radio
resource block) that is chosen 
by user $k$ to transmit the $i$th packet at slot $t$,
denoted by $\ba_k(t;i)$,
to be transmitted at power level\footnote{In this paper,
we assume power-domain NOMA \cite{Ding_CM}
\cite{Choi17_ISWCS}, where multiple signals
in a radio resource block are characterized by  
their (different) power levels.}
$i$ for (power-domain) NOMA mode.
In Section~\ref{S:NOMA}, we discuss 
the power allocation and power levels for opportunistic
NOMA mode in detail.

Throughout the paper, 
for convenience, we assume that $K = M$
and the primary channel of user $k$
is channel $k$, $k = \{1, \ldots, K\}$, for comparisons with OMA.
That is, there are $K$ radio resource blocks for $K$ users and 
the $k$th resource block becomes the primary channel for
user $k$.
Furthermore, we assume that
\be
m (k, i) = [k + i]_K \in \{1, \ldots, K\},
	\label{EQ:mkl}
\ee
where
$[k]_K = \left((k-1)\ {\rm mod}\ K \right) + 1, \ k \in \{1, \ldots, K\}$.
Here, ``${\rm mod}$" represents the modulo operation.
Since there are $K = M$ channels, each user can transmit up to $K$
packets simultaneously. 
However, to avoid the high transmit power in NOMA mode,
we assume that the maximum number of packets to be
simultaneously transmitted is 
limited to $L \ (\le K)$, which is called the depth.
It can be easily shown that as long as $L \le K$,
there is at most one packet at each level in every radio resource
block or channel.
Note that the depth is the number of levels in power-domain NOMA.
In addition, we have OMA if $L = 1$,
i.e., opportunistic NOMA mode with $L= 1$ becomes OMA.

\begin{figure}[thb]
\begin{center}
\includegraphics[width=\figwidth]{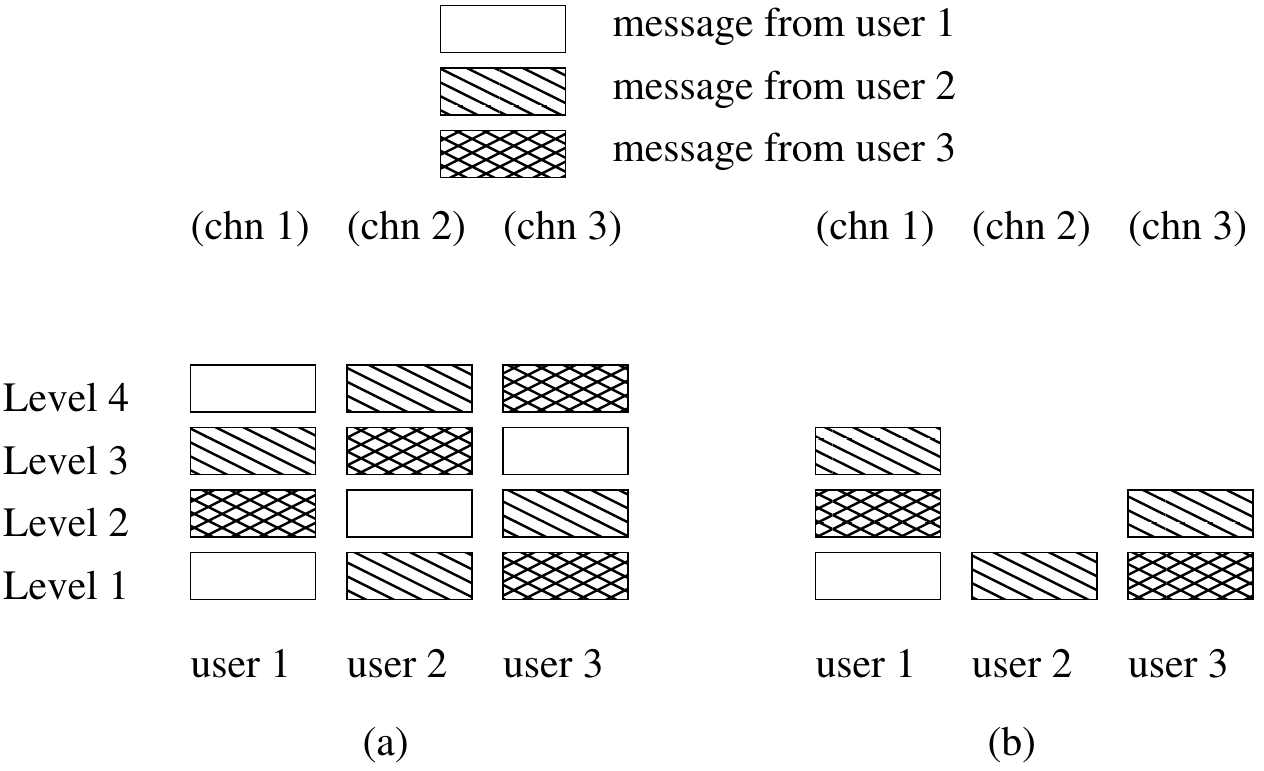}
\end{center}
\caption{Structure of channels to allow NOMA mode:
(a) all possible channels with $K = M = 3$ and depth
$L = 4$;
(b) the used channels if users 1, 2, and 3
have one, three, and two packets to transmit,
respectively.}
        \label{Fig:channels}
\end{figure}

In Fig.~\ref{Fig:channels} (a),
we show the structure of the channels with NOMA
mode when $K = M = 3$ and $L = 4$,
where each pattern is associated to a user's (NOMA) channels\footnote{Note
that each user should be able to transmit packets 
through any of $M$ radio resource blocks.}.
Note that at the channel allocation
in level 4 is the same as level 1 due to the modulo operation 
(with $K = 3$) in \eqref{EQ:mkl}.
Suppose that
each user can have a different number of packets to transmit.
For example, if users 1, 2, and 3
have one, three, and two packets to transmit,
respectively, the channels to be used are as shown in 
Fig.~\ref{Fig:channels} (b).
Note that although more packets (per user)
can be transmitted as $L$ increases,
$L$ cannot be large due to a limited power budget at each user
in power-domain NOMA, where
the transmit power increases with the power level.
Thus, with a modest power budget, $L$ cannot be large
(e.g., $L = 2$).

Denote by $N_k (t)$ the number of packets to be
transmitted from user $k$ at time
$t$, which depends on the CSI and the power budget at user $k$. 
Then, $\bx_{m(k,i)} (t)$ is given by
\be
\bx_{m(k,i)} (t) = 
\left\{
\begin{array}{ll}
h_{m(k,i),k} (t) \ba_k (t;i), & \mbox{if
$i \le N_k (t)$} \cr
0, & \mbox{o.w.} \cr
\end{array}
\right.
\ee
At the BS, the received signal through
channel $m$ is given by
\be
\br_m (t) = \sum_{k = 1}^K 
\sum_{i =1}^{N_k (t)} \bx_{m(k,i)} (t)
\delta_{m, m(k,i)} + \bn_m (t).
	\label{EQ:rss}
\ee

At time slot $t$, user $k$ can transmit $N_k (t)$ packets
using opportunistic NOMA mode.
On the other hand, in OMA, user $k$ can transmit up to one packet
per slot. Therefore, 
if each user has a finite number of packets
to transmit within a certain number of slots
due to a delay constraint,
opportunistic NOMA mode becomes an attractive approach
as it can transmit more than one packet per slot.

\subsection{Asymmetric NOMA System}

In this subsection, we consider a system 
that supports users differently depending on their 
distances from the BS.

\begin{figure}[thb]
\begin{center}
\includegraphics[width=\figwidth]{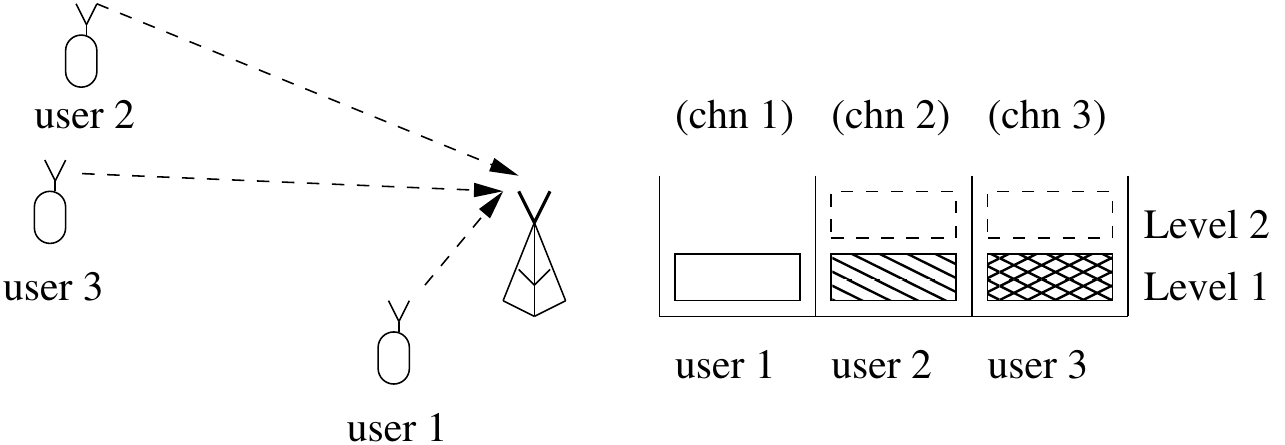}
\end{center}
\caption{An asymmetric system with 3 users with $L = 2$. User 1
is a near user that can employ opportunistic NOMA mode
to transmit another packet through either channel 2 or 3 in level 2,
while users 2 and 3 are far users that are not able to
use opportunistic NOMA mode.}
        \label{Fig:asym}
\end{figure}

Suppose that among $K$ users, 
one user, say user 1, is close to the BS and the other 
$K-1$ users are far away from the BS.
In this case, user 1 can exploit opportunistic NOMA
mode to transmit multiple packets through 
the others' primary channels.
With depth $L = 2$,
to exploit the selection diversity gain if $K-1 > 1$,
it is possible that user 1 can choose one channel from 
channel 2 to channel $K$ 
that has the highest channel gain.
For example, as shown in Fig.~\ref{Fig:asym},
user 1 is able to transmit another packet 
through either channel 2 or 3 in level 2
using opportunistic NOMA mode. On the other hand,
users 2 and 3 can only transmit their packets
through their primary channels\footnote{If 
they have a sufficiently high power budget,
they can employ opportunistic NOMA and transmit packets
through channel 1. Then, this case, where
each user has a sufficient power budget
that can overcome the path loss 
becomes symmetric NOMA.}.

Clearly, there is only one packet transmitted by user 1 in
channel 1. On the other hand,
there can be two packets in
the received signal through channel $k$
as follows:
\be
\br_k (t) = h_{k,k}(t) \ba_k (t;1) + h_{k,1}(t) \ba_1 (t;2) + \bn_k (t),
	\label{EQ:ark}
\ee
if channel $k \in \{2, \ldots, K\}$ is chosen by user 1 to transmit
an additional packet.
In the asymmetric system, user 1 (i.e., a near user)
can take advantage of a high channel gain for opportunistic NOMA.
To see this, we can consider the SINR
of user 1 (i.e., near user) in channel $k$,
denoted by $\gamma_{1,k}$, as follows:
\be
\gamma_{1,k} = \frac{|h_{k,1} (t)|^2 P_{1;2} (t) }{
|h_{k,k} (t)|^2 P_{k;1} (t) + N_0 },
\ k \in \{2,\ldots, K\},
\ee
where $P_{k;l} (t)$ is the transmit power of user $k$'s packet
in level $l$ at time $t$.
Since user $k \in \{2, \ldots, K\}$ is a far user,
we expect that
$\uE[|h_{k,1}(t)|^2 ] \gg \uE[|h_{k,k}(t)|^2 ]$, 
$k \in \{2, \ldots, K\}$.
Thus, the resulting SINR can be high without having a high
transmit power of user 1's packet in level 2. In other words,
user 1 can employ 
opportunistic NOMA mode without a high transmit power
thanks to the difference propagation loss 
between near and far users.
In addition, if $K > 2$, user 1 can choose the
channel that has the highest gain among $K-1$ others' channels,
i.e., $\max_{2 \le k \le K} |h_{k,1}(t)|^2$,
which provides a (selection) diversity gain.
The resulting case is referred to as the selection diversity
based opportunistic NOMA (SDO-NOMA) mode.

Alternatively, it is possible to transmit
up to $K-1$ additional packets in level 2
in the asymmetric system.
In this case, $K-1$ additional packets
from user 1 can be received at slots $k$, $k = 2,
\ldots, K$, as follows:
\be
\br_k (t) = h_{k,k}(t) \ba_k (t;1) + h_{k,1}(t) \ba_1 (t;k) + \bn_k (t),
\ee
for all $k \in \{2, \ldots, K\}$.
The resulting case is referred to as fully opportunistic
NOMA (FO-NOMA) mode.

It can be seen that SDO-NOMA 
exploits the others' channels to transmit
one additional packet using the selection diversity gain,
while FO-NOMA can transmit up to $K-1$ additional 
packets through the others' channels.
Thus, at the BS, one SIC is required in SDO-NOMA, but multiple
SICs are required for FO-NOMA.
This implies that FO-NOMA can transmit more packets
than SDO-NOMA at the cost of 
a higher receiver complexity.

\section{Power Allocation for Opportunistic NOMA Mode}
\label{S:NOMA}

In this section, we discuss the power allocation
when opportunistic NOMA mode is used with a limited power budget.

To allow SIC at the BS, we assume that
each level has a target or required SINR
and the power allocation is performed to meet
the target SINR.
Let $\rho_l$ represent the received signal power
at level $l$. Then, provided that SIC is successful
to decode the signals in levels $l+1, \ldots, L$,
the SINR of the packet in level $l$ becomes
\be
\gamma_l = \frac{\rho_l}{\sum_{m=1}^{l-1}  \rho_m + N_0},
\ l = 1, \ldots, L.
	\label{EQ:gamme}
\ee
For simplicity,
we assume that all the packets are encoded at the same rate. Thus,
the required SINR 
for successful decoding, denoted by $\Gamma$,
becomes the same for all
levels,
i.e., $\gamma_l = \Gamma$, $l = 1,\ldots, L$.
Thus, from \eqref{EQ:gamme},
$\{\rho_l\}$ can be recursively decided as follows:
\be
\rho_l = \Gamma \left(
\sum_{m=1}^{l-1}  \rho_m + N_0 \right).
	\label{EQ:rr}
\ee
Provided $\Gamma \ge 1$, we can see that $\rho_l$ increases
with $l$.

\subsection{Symmetric System}

Consider user 1 at time slot $t$. 
For convenience, we omit the time index $t$ and user index $k$. 
In this case, $h_m$ represents $h_{m,1} (t)$,
i.e., $h_m = h_{m,1} (t)$.
In the symmetric system,
when user 1 transmits a packet through channel $m$,
its power level is also $m$ according to \eqref{EQ:mkl}.
Thus, the transmit power
of the packet of user 1 to be transmitted 
through channel $m$ (and level $m$),
denoted by $P_m$, is decided to satisfy
\be
|h_m|^2 P_m = \rho_m, \ m = 1, \ldots, K.
	\label{EQ:hPb}
\ee
Since each user has a limited power budget to transmit packets
in each slot, denoted by $\Omega$,
the maximum number of the packets per slot time interval
in the symmetric system becomes
\be
N^* = \min \left\{ \max_N \left\{N \,|\, \sum_{m=1}^N P_m \le \Omega 
\right\}, L\right\},
	\label{EQ:n_s}
\ee
where $P_m = \frac{\rho_m}{|h_m|^2}$
from \eqref{EQ:hPb}.
Here, $N^*$ becomes $N_k (t)$ in \eqref{EQ:rss},
while we have OMA if $L = 1$.


\subsection{Asymmetric System}

In the asymmetric system,
only user 1 can employ opportunistic NOMA mode
to allow to transmit more than one packet per time slot.
Thus, for user $k \in \{2,\ldots,K\}$,
it follows
$$
|h_{k,k}|^2 P_k = \rho_1,
$$
and for user 1, the power allocation is performed to hold
\begin{align}
|h_{1,1}|^2 P_{1,1} = \rho_1 \ \mbox{and} \
\max_{2 \le k \le K} |h_{k,1}|^2 P_{k,1} = \rho_2
\end{align}
in SDO-NOMA mode. As a result, user 1 can transmit two packets if
\be
\frac{\rho_1}{|h_{1,1}|^2} +
\frac{\rho_2}{\max_{2 \le k \le K} |h_{k,1}|^2} \le \Omega.
	\label{EQ:sdo}
\ee

In FO-NOMA mode, user 1 can transmit (at least) $n$ packets if
\be
\frac{\rho_1}{|h_{1,1}|^2} +
\sum_{m=1}^{n-1} \frac{\rho_2}{ g_{(m)} } \le \Omega,
\ee
where $g_{(m)}$ denotes the $m$th largest order statistic of
$|h_{2,1}|^2, \ldots, |h_{K,1}|^2$.

\subsection{Some Issues}

In this paper, we assume that
each user has perfect CSI so that the power allocation can
be performed to hold \eqref{EQ:hPb}.
However, due to the background noise,
a user only has an estimate of CSI, which may lead to
imprecise power allocation and the resulting SINR 
can be different from the target SINR, $\Gamma$.
Due to the different SINR from the target SINR,
decoding failure and erroneous SIC become inevitable 
\cite{Liu18} \cite{Cui18}.
Thus, in order to avoid them due to 
imprecise power allocation, 
a margin can be given to the target SINR, $\Gamma$.

It is also assumed that if the SINR is greater than
or equal to $\Gamma$, the BS is able to decode packets
\cite{Choi14} \cite{Ding14} \cite{Ding_CM}.
If the length of packet is sufficiently long
and capacity-achieving
codes are used, $\Gamma$ can be decided to satisfy
$R < \log_2 (1+ \Gamma)$,
where $R$ is the transmission or code rate \cite{CoverBook}
\cite{MacKayBook}.
However, as shown in \cite{Polyanskiy10IT} \cite{Durisi16},
for short packets, it is necessary to take into account
the channel dispersion, which makes the required
SINR higher than $2^{R} - 1$.
In particular, from \cite{Durisi16},
the achievable rate for a finite-length code
is given by
$$
R(n,\epsilon) \approx \log_2 (1+ \Gamma) -
\sqrt\frac{V(\Gamma)}{n} \cQ^{-1} (\epsilon) + O \left(
\frac{\log_2 n}{n}
\right) ,
$$
where $n$ represents the length of codeword for each packet
and $\epsilon$ is the error probability.
With a sufficiently low $\epsilon$, for given $R$ and $n$, we can set
$\Gamma$ to satisfy $R \le R(n, \epsilon)$.
It is noteworthy that $\epsilon$ cannot be zero
with short packets. 
Thus, in deciding $\Gamma$ as above, $\epsilon$ has to be
sufficiently low 
and negligible
compared to the target session error probability
(the session error probability
will be discussed in Section~\ref{S:OA}).


\section{Probability of Multi-Packet Transmissions When $L = 2$} \label{S:PMP}

In this section, we find the 
probability of multi-packet transmissions
using opportunistic NOMA mode. For tractable analysis,
we only consider the case of $L = 2$.

\subsection{Symmetric System}

In the symmetric system,
each user can equally employ 
opportunistic NOMA mode. Thus, 
it might be reasonable to assume that the statistical channel
conditions of the users are similar to each other.
In particular, in this subsection, we assume that
$h_{m,k}$ is 
independent and identically distributed (iid) for tractable analysis
and focus on finding
the probability of multi-packet transmissions for
iid channels.
Thanks to the symmetric channel
condition for all users (as $h_{m,k}$ is iid),
it is sufficient to consider
one user and let $h_m = h_{m,k}$ (i.e., omitting the user index, $k$)
in this subsection.

According to \eqref{EQ:n_s},
the probability that a user can transmit at least $m$ packets is given by
\be
\beta_m = \Pr \left(\sum_{i=1}^m \frac{\rho_i}{|h_i|^2} 
\le \Omega \right).
\ee
Denote by  $\alpha_m$
the probability that user 1 can transmit $m$ packets.
That is, $\alpha_m = \Pr(M_{\rm succ} = m)$, while
$\beta_m = \Pr(M_{\rm succ} \ge m)$,
where $M_{\rm succ}$ 
denotes the number of successfully transmitted packets.
For OMA (i.e., $L = 1$),
it follows
$$
\alpha_0 = 1 - \beta_1 \ \mbox{and} \ \alpha_1 = \beta_1,
$$
while $\alpha_m = 0$ for all $m \ge 2$.

In opportunistic NOMA mode with $L = 2$, we only need to consider
$\beta_1$ and $\beta_2$ to find $\alpha_m$, $m \in \{0,1,2\}$.
Clearly,
$\alpha_0 = 1 - \beta_1$, 
$\alpha_1 = \beta_1 - \beta_2$, and $\alpha_2 = \beta_2$.

Note that the $\beta_m$'s are independent of 
the depth $L$, which is also true for  
$\alpha_m = \beta_m - \beta_{m+1}$, $m = 0, \ldots, L-1$
(with $\beta_0 = 1$).
However, $\alpha_m$, $m = L$, depends on $L$ as shown above
(while $\alpha_m = 0$ for $m > L$).
Thus, in order to emphasize it, with a finite $L$,
the probability of that user 1 can transmit $L$ packets
is denoted by $\bar \alpha_L$ instead of $\alpha_L$.

\begin{mylemma}
Suppose that $|h_{m}|^2$ is 
iid and has
an exponential distribution, i.e.,
$|h_{m}|^2 \sim {\rm Exp} (1)$.
That is, independent Rayleigh fading channels are assumed.
Then, we have
\begin{align}
\beta_1 & = e^{- \frac{\rho_1}{\Omega}} \cr
\beta_2 & = e^{- \frac{\rho_1+ \rho_2}{\Omega}}
\frac{2 \sqrt{\rho_1 \rho_2}}{\Omega}
{\rm K}_1 \left( \frac{2 \sqrt{\rho_1 \rho_2}}{\Omega}
\right),
	\label{EQ:L1}
\end{align}
where ${\rm K}_\nu (x)$ is the 
modified Bessel function of the second kind 
which is given by
${\rm K}_\nu (x)
= \int_0^\infty e^{-x \cosh t} \cosh(\nu t) d t$.
\end{mylemma}
\begin{IEEEproof}
It can be shown that
\begin{align}
\beta_1 & = \Pr \left( \frac{\rho_1}{|h_{1}|^2} \le \Omega \right) \cr
& = \Pr \left( |h_{1}|^2 \ge \frac{\rho_1}{\Omega} \right) 
= e^{- \frac{\rho_1}{\Omega}}.
\end{align}

Letting $X_1 = |h_{1}|^2$ and
$X_2 = |h_{2}|^2$,
it can also be shown that
\begin{align}
\beta_2 & = 
\Pr \left( \frac{\rho_1}{X_1} + \frac{\rho_2}{X_2} \le \Omega \right) \cr
& = \Pr \left(  \frac{\rho_2}{X_2} \le \Omega 
-  \frac{\rho_1}{X_1} \right) \cr
& = \int_{\frac{\rho_1}{\Omega}}^\infty 
\exp \left( - \frac{\rho_2 x_1}{\Omega x_1 - \rho_1} \right)
 e^{- x_1} d x_1 \cr
& = \int_0^\infty 
\exp \left( - \frac{\rho_2}{t_1} \frac{t_1+ \rho_1}{\Omega} \right)
e^{- \frac{t_1 + \rho_1}{\Omega}} d t_1 \cr
& = 
e^{- \frac{\rho_1+ \rho_2}{\Omega}}
\int_0^\infty
\exp \left( - \frac{\rho_1 \rho_2}{\Omega} \frac{1}{t_1} 
- \frac{t_1}{\Omega} \right) d t_1,
	\label{EQ:ab2}
\end{align}
where $t_1 = \Omega x_1 - \rho_1$.
After some manipulations, 
we can further show that
\be
\int_0^\infty
\exp \left( - \frac{\rho_1 \rho_2}{\Omega} \frac{1}{t_1} 
- \frac{t_1}{\Omega} \right) d t_1
= \frac{2 \sqrt{\rho_1 \rho_2}}{\Omega}
{\rm K}_1 \left( \frac{2 \sqrt{\rho_1 \rho_2}}{\Omega}
\right).
	\label{EQ:aK}
\ee
Substituting \eqref{EQ:aK} into \eqref{EQ:ab2},
we can obtain the expression for $\rho_2$ in \eqref{EQ:L1},
which completes the proof.
\end{IEEEproof}

\subsection{Asymmetric System}

Unlike the symmetric system, when the asymmetric system
is considered, it is expected that
the long-term channel gain of user 1 is greater than those of the others.
Thus, under Rayleigh fading, we assume that 
\begin{align*}
h_{k,1} & = \sqrt{\varphi_1} U_{k,1}, \ k \in \{1,\ldots,K\} \cr
h_{k,k} & = \sqrt{\varphi_k} U_{k,k}, \ k \in \{2,\ldots,K\},
\end{align*}
where $U_{m,k}$ is an independent zero-mean CSCG random variable
with unit variance, i.e., $U_{m,k} \sim \cC \cN(0,1)$,
which represents the short-term fading coefficient 
for $h_{m,k}$. Furthermore, $\varphi_{k}$
denotes the long-term fading coefficient for user $k$
(so that $\uE[|h_{m,k}|^2] = \varphi_k$).
In general, the long-term channel coefficient
of a user is decided by the distance between
the BS and the user \cite{TseBook05}.
Then, letting $d_{k}$ denote the distance
between the BS and user $k$, 
if the long-term channel gain of user 1 is normalized,
it can be shown that
$\varphi_k = \left(\frac{d_1}{d_k} \right)^\zeta$,
where $\zeta$ represents the
path loss exponent.
Consequently, if we assume that $\varphi_k = \sigma^2$ for
$k = 2, \ldots, K$, it can be shown that
\begin{align}
|h_{k,1}|^2 & \sim {\rm Exp}(1),\ k \in \{1, \ldots, K\} \cr
|h_{k,k}|^2 & \sim {\rm Exp}(\sigma^2), \ k \in \{2, \ldots, K\},
	\label{EQ:hha}
\end{align}
where $\sigma^2 \ll 1$, which will be used for the analysis in this subsection.

\begin{mylemma}
Assume that the channel coefficients are given as in \eqref{EQ:hha}.
Suppose that far users 
(e.g., users $2, \ldots, K$) have the power budget, $\bar \Omega$.
Then, the probability of transmission through
primary channel from user $k
\in \{2,\ldots, K\}$, denoted by $\beta_{1,k}$, is given by
\begin{align}
\beta_{1,k} = 
\exp
\left(- \frac{\rho_1}{\sigma^2 \bar \Omega}\right), \ k \in \{2,\ldots,K\}.
	\label{EQ:L2a}
\end{align}
In SDO-NOMA mode,
the probability that 
user 1 can transmit at least $n \in \{1,2\}$ packets with power budget $\Omega$,
denoted by $\beta_{n,1}$, is given by
\be
\beta_{1,1} = e^{- \frac{\rho_1}{\Omega}}
	\label{EQ:L2b}
\ee
and
\begin{align}
\beta_{2,1} 
& = 
\sum_{m=1}^{K-1} \binom{K-1}{m}
(-1)^{m+1} e^{- \frac{\rho_1 + m \rho_2}{\Omega}} \cr
& \quad \times
\frac{2 \sqrt{m \rho_1 \rho_2}}{\Omega}
{\rm K}_1
\left(\frac{2 \sqrt{m \rho_1 \rho_2}}{\Omega} \right).
	\label{EQ:L2c}
\end{align}
\end{mylemma}
\begin{IEEEproof}
Since the derivations of \eqref{EQ:L2a}
and \eqref{EQ:L2b}
are straightforward, we omit them.

From \eqref{EQ:sdo},
we have
\be
\beta_{2,1}= \Pr \left(
\frac{\rho_1}{X_1} + \frac{\rho_2}{Z} 
\le \Omega \right),
	\label{EQ:sdo_1}
\ee
where $X_1 = |h_{1,1}|^2$ and $Z = \max_{2 \le k \le K} |h_{k,1}|^2$.
Under \eqref{EQ:hha}, since the 
cumulative distribution function (cdf) of 
the order statistic $Z$ \cite{DavidBook} is given by
$F_Z (z) = \left(
1 - e^{-z}
\right)^{K-1}$,
it can be shown that
\begin{align}
\beta_{2,1} & = 
\Pr \left(  \frac{\rho_2}{Z} \le \Omega 
-  \frac{\rho_1}{X_1} \right) \cr
& = \int_{\frac{\rho_1}{\Omega}}^\infty 
1 - \left(1-
e^{ - \frac{\rho_2 x_1}{\Omega x_1 - \rho_1}} \right)^{K-1}
 e^{- x_1} d x_1 \cr
& = e^{- \frac{\rho_1}{\Omega}} - 
\int_{\frac{\rho_1}{\Omega}}^\infty 
\sum_{m=0}^{K-1} \binom{K-1}{m}
\left(- e^{ - \frac{\rho_2 x_1}{\Omega x_1 - \rho_1}} \right)^{m} 
 e^{- x_1} d x_1 \cr
& = 
e^{- \frac{\rho_1}{\Omega}} \cr
& - \sum_{m=0}^{K-1} \binom{K-1}{m}
(-1)^m
\int_{\frac{\rho_1}{\Omega}}^\infty 
e^{ - \frac{m \rho_2 x_1}{\Omega x_1 - \rho_1}} e^{- x_1} d x_1.
	\label{EQ:L2_n}
\end{align}
As in \eqref{EQ:aK}, we can show that
\be
\int_{\frac{\rho_1}{\Omega}}^\infty 
e^{ - \frac{m \rho_2 x_1}{\Omega x_1 - \rho_1}} e^{- x_1} d x_1
=
\frac{2 \sqrt{m \rho_1 \rho_2}}{\Omega}
{\rm K}_1
\left(\frac{2 \sqrt{m \rho_1 \rho_2}}{\Omega} \right).
	\label{EQ:mK}
\ee
Substituting \eqref{EQ:mK} into \eqref{EQ:L2_n},
we have
\begin{align}
\beta_{2,1} & = e^{- \frac{\rho_1}{\Omega}}  - \sum_{m=0}^{K-1} \binom{K-1}{m}
(-1)^m e^{- \frac{\rho_1 + m \rho_2}{\Omega}} \cr
& \quad \times
\frac{2 \sqrt{m \rho_1 \rho_2}}{\Omega}
{\rm K}_1
\left(\frac{2 \sqrt{m \rho_1 \rho_2}}{\Omega} \right),
\end{align}
which is identical to \eqref{EQ:L2c}. This completes the proof.
\end{IEEEproof}

Note that $\beta_{1,1}$ and $\beta_{2,1}$
are independent of $\varphi_k$ or $\sigma^2$ as they
are decided by the channel gains of user 1,
i.e., $|h_{k,1}|^2$.

\subsection{The Case of $L > 2$}

In this section, 
as mentioned earlier,
we mainly focus on the case of depth 2,
i.e., $L = 2$. In general, 
if $L > 2$, it is difficult to obtain closed-form expressions 
for the $\alpha_m$'s (or $\beta_m$'s).
However, we can show that the case of $L = 2$
provides the worst performance of opportunistic NOMA mode
with $L \ge 2$.
In particular, with the average number of 
transmitted packets per slot, we have the following result.

\begin{mylemma}	\label{L:bN}
With $L \ge 2$ in opportunistic NOMA mode,
let $\bar N_L$ be the 
average number of 
transmitted packets per slot.
Then, $\bar N_L$ increases in $L$, i.e.,
\be
\bar N_2 \le \bar N_3 \le  \ldots.
\ee
\end{mylemma}
\begin{IEEEproof}
With $L \ge 2$, let $\bar \alpha_L = \beta_L$. 
Consider the case of $L = 2$, where have
$\alpha_0 = 1 - \beta_1$, $\alpha_1 = \beta_1 - \beta_2$, 
and $\bar \alpha_2 = \beta_2$.
Note that when $L = 3$, we have $\alpha_2 = \beta_2 - \beta_3$
and $\bar \alpha_3 = \beta_3$.
It can be shown that
\begin{align*}
\bar N_2 = \alpha_1 + 2 \bar \alpha_2 \ \mbox{and} \
\bar N_3 = \alpha_1 +  2 \alpha_2 +3 \bar \alpha_3.
\end{align*}
From this, it follows that
$\bar N_3 - \bar N_2 = \bar \alpha_3 \ge 0$,
because $\bar \alpha_3$ is a probability.
Similarly,
we can also show that 
$\bar N_{L+1} - \bar N_L = \bar \alpha_{L+1} \ge 0$,
which completes the proof.
\end{IEEEproof}

Consequently, throughout the paper, for opportunistic NOMA mode,
we only consider the case of $L = 2$ for analysis,
which can be used as 
performance bounds for the case of $L > 2$.

\section{Session Error Analysis}	\label{S:OA}

In this section,
it is assumed that each user has a set of $W$ packets, 
which is called a stream, to be delivered to the BS
within a finite time.
If a user can transmit one packet during every slot, 
the transmission of 
a stream can be completed within $W$ slots.
However, due to deep fading,
some packets are to be re-transmitted,
which requires additional time slots.
As a result,
we may need to have $W_{\rm S}\ (\ge W)$ slots for the
transmission of a stream. 
For convenience, $W_{\rm S}$ is called the length of session\footnote{It
is assumed that one session is required
to transmit $W$ packets or a stream,
which is to be completed within a time duration of $W_{\rm S}$ slots.}.
Clearly, it is expected to design
a system for SMDDC that can complete 
the delivery of a stream within 
a session time (corresponding to the time
period of $W_{\rm S}$ slots) 
with a high probability.
In this section,
we discuss the session error probability,
which is the probability that a stream cannot be delivered
within a session time, in terms of $W$ and $W_{\rm S}$
(the corresponding event is referred to as
a session error).

Note that after a session,
the BS needs to send a feedback signal if 
a session error happens. In addition, the
BS can send the indices of unsuccessfully decoded packets
among $W$ packets so that a user can re-transmit them.
Since a session error event results in a long delay, it is necessary
to keep it low for low-delay transmissions.

\subsection{An Upper-bound on Session Error Probability}

Let $D(t) = \min\{t, W\} - S(t)$, where $S(t)$
denotes the accumulated number of successfully transmitted
packets at time $t \in \{1, \ldots, W_{\rm S}\}$.
In addition, denote by $V(t)$ the number of successfully
transmitted packets at time slot $t$.
Note that $V(t) \in \{0,1\}$ in OMA and $\in \{0,1,2\}$
in opportunistic NOMA.
Clearly, $S(t) = \sum_{i=1}^t V(i)$.
If there are any packets that cannot
be successfully transmitted due to fading in the first $W$ slots, 
we may have $\sum_{i=1}^W V(i) < W$, which results in
$D(W) > 0$. Thus, when $D(W)$ is positive,
more slots are needed to transmit unsuccessful packets.
If $D(t) = 0$ for any $t \in \{W, \ldots, W_{\rm S}\}$,
we can see that all $W$ packets can be successfully
transmitted within $W_{\rm S}$ slots.
From this, $D(t)$, $t \in \{W, \ldots, W_{\rm S}\}$, can be used 
to see successful transmission of a short-message 
or stream (i.e., a set of $W$ packets)
within a given time (i.e., 
a total duration of $W_{\rm S}$ slots).
It can be shown that
\be
D(t) = 
\left\{ 
\begin{array}{ll}
D(t-1) + 1 - V(t), & \mbox{if $t \le W$} \cr
D(t-1) - V(t), & \mbox{if $W < t \le W_{\rm S}$} \cr
\end{array}
\right.
	\label{EQ:Dt}
\ee
where $D(0) = 0$.
Thus, $D(t)$ becomes a Markov chain with
the following transition probability:
\begin{align}
\Pr( D(t) = d+ \indicator(t \le W)- n \,|\, D(t-1) = d) = \alpha_n,
\end{align}
where $\indicator(\cA)$ 
represents the indicator function of event $\cA$.

A session error event happens 
if there are packets that are not yet transmitted
after $W_{\rm S}$ uses of channel.
Thus, using $D(W_{\rm S})$, the session error probability 
can be expressed as
\begin{align}
P_{\rm SE} (W_{\rm S}) 
= \Pr (D(W_{\rm S})> 0) 
= \Pr\left( \sum_{t=1}^{W_{\rm S}} V(t)  
< W \right).
	\label{EQ:PPW}
\end{align}
In order to have a low session error probability,
it is necessary to hold
$W_{\rm S} \uE[V(t)] > W$
or
\be
\uE[V(t)] > \frac{W}{W_{\rm S}} \deft \kappa.
	\label{EQ:NS}
\ee
For convenience, define the relative delay as 
$\tau_\kappa = \frac{W_{\rm S}}{W} = \frac{1}{\kappa}$.
In OMA, it is expected to have
$\uE[V(t)] \le 1$. Thus, $\kappa$ cannot be close to $1$,
which means a long relative delay.
On the other hand, if opportunistic NOMA mode is used,
we can have $\uE[V(t)] \ge 1$ (as more than one packet
can be transmitted within a slot).
In this case, $\kappa$ can be close 1 (or even greater than
1). Thus, 
a short relative delay can be achieved (with a low 
session error probability).
This clearly demonstrates the advantage of opportunistic NOMA
mode over OMA for SMDDC.

In general, the session
error probability decreases as $W_{\rm S}$ increases
or $\kappa$ decreases.
However, a small $\kappa$
or a large $W_{\rm S}$ is not desirable for SMDDC.
Therefore, it is 
necessary to decide a minimum $W_{\rm S}$ with a certain
target session error probability.
To this end, we need to have 
a closed-form expression
for the session error probability in terms of key parameters
including $W$ and $W_{\rm S}$.
However, since it is not easy to find an exact expression,
we resort to an upper bound using the Chernoff bound \cite{Mitz05}. 

For an upper-bound on
the session error probability, 
from \eqref{EQ:PPW}, we consider
the following inequality:
\begin{align}
P_{\rm SE} (L) 
& \le \bar P_{\rm SE} (L, \lambda)  \cr
& = e^{\lambda W} \left(
\uE[e^{-\lambda V(t)}] \right)^{W_{\rm S}}, \ \lambda > 0.
	\label{EQ:UB}
\end{align}
The Chernoff bound is given by
\be
\bar P_{\rm SE} (L) = \min_{\lambda > 0} \bar P_{\rm SE} (L, \lambda).
	\label{EQ:CB}
\ee
Here, $\lambda^*$ that minimizes $\bar P_{\rm SE} (L, \lambda)$
is given by
\be
\lambda^* = \argmin_{\lambda \ge 0} 
\bar P_{\rm SE} (L, \lambda)
 = \argmin_{\lambda \ge 0} 
e^{\lambda W} \left(
\uE[e^{-\lambda V(t)}] \right)^{W_{\rm S}}.
	\label{EQ:lam_s}
\ee

\subsection{The Case of OMA}

\begin{mylemma}	\label{L:4}
In OMA (i.e., with $L = 1$), 
the Chernoff bound on the session error
probability\footnote{Since $V(t) 
\in \{0,1\}$, the session error probability 
can be given by $\sum_{w =0}^{W-1} \binom{W_{\rm S}}{w} \bar \alpha_1^w
\alpha_0^{W_{\rm S} - w}$. Thus,
the upper-bound in \eqref{EQ:CB_L1}
can be found from the binomial distribution,
which is a well-known result \cite{Arratia89}.}
is given by
\begin{align}
\bar P_{\rm SE} (1)
\le 
\left[\left( \frac{\bar \alpha_1}{\kappa} \right)^\kappa
\left( \frac{1-\bar \alpha_1}{1-\kappa} \right)^{1-\kappa} \right]^{W_{\rm S}}.
	\label{EQ:CB_L1}
\end{align}
Here, it is necessary that
$\bar \alpha_1 > \kappa$
for the condition \eqref{EQ:NS} since $\uE[V(t)] = \bar \alpha_1$.
\end{mylemma}
\begin{IEEEproof}
With $L = 1$, since
$\uE[e^{-\lambda V(t)}] = \alpha_0 + \bar \alpha_1 e^{-\lambda}$,
we have
\begin{align}
\ln \left(
e^{\lambda W} \left(
\uE[e^{-\lambda V(t)}] \right)^{W_{\rm S}} \right)
& = \lambda W + W_{\rm S} \ln
\left(\alpha_0 + \bar \alpha_1 e^{-\lambda} \right).
\end{align}
Taking the differentiation with respect to $\lambda$
and setting it to zero (since the upper-bound
is convex in $\lambda$ \cite{Dembo98}), 
it can be shown that
\be
e^{-\lambda^*} = \frac{\kappa}{1- \kappa} \frac{\alpha_0}{\bar \alpha_1}.
	\label{EQ:L3e}
\ee
Note that if $\bar \alpha_1 > \kappa$ (for the necessary
condition in \eqref{EQ:NS}),
we can show that $\lambda^* > 0$.

Substituting \eqref{EQ:L3e} into \eqref{EQ:CB},
we can have \eqref{EQ:CB_L1},
which completes the proof.
\end{IEEEproof}

Note that using the weighted 
arithmetic mean (AM) and geometric mean (GM) inequality \cite{Bullen13},
we can show that
$$
\left( \frac{\bar \alpha_1}{\kappa} \right)^\kappa
\left( \frac{1-\bar \alpha_1}{1-\kappa} \right)^{1-\kappa} \le 
\kappa
\left( \frac{\bar \alpha_1}{\kappa} \right) +
(1-\kappa) \left( \frac{1-\bar \alpha_1}{1-\kappa} \right) = 
1,
$$
which implies
that the upper-bound in \eqref{EQ:CB_L1} cannot be greater than 1.

According to \eqref{EQ:NS}, the minimum achievable
relative delay,
$\tau_\kappa$, becomes  $\frac{1}{\bar \alpha_1}$ in OMA, 
which can be achieved as $\kappa \to \bar \alpha_1$.
However, in this case,
the session error probability can be high
since 
$\left( \frac{\bar \alpha_1}{\kappa} \right)^\kappa
\left( \frac{1-\bar \alpha_1}{1-\kappa} \right)^{1-\kappa} \to 1$
as  $\kappa \to \bar \alpha_1$ for a finite $W_{\rm S}$.
Thus, we need $\kappa \ll \bar \alpha_1 < 1$ for 
a low session error probability, which implies a long relative delay.
In other words, OMA is not suitable for SMDDC.

\subsection{The Case of Opportunistic NOMA}

Using the Chernoff bound in \eqref{EQ:CB}, we can find an upper-bound
on the session error probability 
when opportunistic NOMA mode is employed
as follows.

\begin{mylemma}	\label{L:5}
In opportunistic NOMA mode with $L = 2$, 
if $\uE[V(t)] = \alpha_1 + 2 \bar \alpha_2 > \kappa$,
the Chernoff bound is given by
\be
\bar P_{\rm SE} (2)
\le 
\left(e^{\kappa \lambda^*} (
\alpha_0 + \alpha_1 e^{-\lambda^*}
+ \bar \alpha_2 e^{-2 \lambda^*})\right)^{W_{\rm S}},
	\label{EQ:CB_L2}
\ee
where
\be
e^{-\lambda^*} = \frac{\sqrt{(1-\kappa)^2 \alpha_1^2 + 4 \kappa
(2 - \kappa) 
\alpha_0 \bar \alpha_2} - (1-\kappa) \alpha_1}{2 (2 - \kappa) \bar \alpha_2}.
\ee
\end{mylemma}
\begin{IEEEproof}
Since the proof is similar to that of Lemma~\ref{L:4}, we
omit it.
\end{IEEEproof}

Although we can 
obtain the session error probabilities of OMA and
opportunistic NOMA  mode from \eqref{EQ:CB_L1} and \eqref{EQ:CB_L2}
using upper-bounds, respectively,
it is difficult to directly see
the gain by using opportunistic NOMA mode
for SMDDC.
Thus, based on the upper-bound in \eqref{EQ:UB},
we consider the NOMA factor that is given by
\begin{align}
\eta 
= \min_{\lambda > 0} 
\left(\frac{ \bar P_{\rm SE} (2, \lambda)}{ \bar P_{\rm SE} (1, \lambda)} 
\right)^{1/W_{\rm S}}.
	\label{EQ:eta_0}
\end{align}
It can be seen that for a given length of session $W_{\rm S}$, 
$\eta^{W_{\rm S}}$ becomes the minimum possible ratio
of the session error probability of opportunistic NOMA mode to that of OMA.
Note that it is desirable that the NOMA factor, $\eta$, 
is being independent of the values of $W$ and $W_{\rm S}$ 
so that $\eta$ can demonstrate the pure gain of opportunistic NOMA.

\begin{mylemma}
The NOMA factor is given by
\be
\eta = 1 - \frac{\bar \alpha_2}{(1+ \sqrt{\alpha_0})^2}.
	\label{EQ:eta}
\ee
\end{mylemma}

\begin{IEEEproof}
From \eqref{EQ:eta_0} and \eqref{EQ:UB}, we can show that
\begin{align}
\eta 
= 
\min_{0 \le z < 1}
\frac{\alpha_0 + \alpha_1 z + \bar \alpha_2 z^2}
{\alpha_0 + (\alpha_1 + \bar \alpha_2) z},
	\label{EQ:eta_a1}
\end{align}
where $z = e^{-\lambda}$.
Clearly, \eqref{EQ:eta_a1}
is a fractional program \cite{Schaible81},
where the numerator is convex and the denominator is concave
in $z$. In particular, it is a convex-concave fractional program,
which can be reduced to a convex program \cite{Schaible81}.
Thus, the (unique) solution can be found by taking
the differentiation with respect to $z$
and setting it to zero. Then, the optimal $z$,
which is denoted by $z^*$, needs to satisfy
the following equation:
\be
\bar \alpha_2 (1- \alpha_0) z^2 + 2 \bar \alpha_2 \alpha_0 z
-\bar \alpha_2 \alpha_0 = 0.
\ee
After some manipulations, we have
\be
z^* = \frac{\sqrt{\alpha_0}}{1+ \sqrt{\alpha_0}} < 1.
	\label{EQ:z_s}
\ee
Substituting \eqref{EQ:z_s}
into \eqref{EQ:eta_a1},
we can have \eqref{EQ:eta}, which completes the proof.
\end{IEEEproof}

From \eqref{EQ:eta}, we can see that 
the NOMA factor, $\eta$, decreases with $\bar \alpha_2 = \beta_2$
and increases with $\alpha_0$.
In addition, as long as $\bar \alpha_2 > 0$, $\eta$
becomes less than 1.
From this, with $\eta < 1$, it is expected that the 
session error probability will be dramatically lowered by 
opportunistic NOMA mode compared to OMA
for a reasonably long session length, $W_{\rm S}$.
For example, 
assuming that the upper-bound on the session error probability of OMA is 1,
if $\eta = 1 - \epsilon \approx e^{-\epsilon}$ (for $\epsilon \ll 1$),
the session error probability of
opportunistic NOMA mode 
becomes $e^{-\epsilon W_{\rm S}}$
(note that it might be a lower-bound
as $\eta^{W_{\rm S}}$ is the minimum possible ratio
of session error probabilities). 
In particular, if $(\epsilon, W_{\rm S}) = (0.1, 50)$,
the session error probability of
opportunistic NOMA mode 
can be as low as $e^{-5} = 0.0067$.

\section{Simulation Results}	\label{S:Sim}

In this section, we present simulation
results for user 1's performance under
the assumption that the channels
of $K = M$ radio resource blocks 
experience
independent Rayleigh fading, i.e., $h_{m,1} \sim \cC \cN(0,1)$ or
$|h_{m,1}|^2 \sim {\rm Exp}(1)$, $m = 1, \ldots, K$.
For convenience, we also assume that $N_0 = 1$.

Note that we mainly consider the performance of user 1 in this section
for the following reasons. 
In symmetric NOMA, due to symmetric conditions,
the performance of user 1 is the same as that of another user.
In asymmetric NOMA, user 1, i.e., the strong
user, is only the user employing opportunistic NOMA, 
while the other users can be seen as
users in conventional OMA (as a result,
their performance is identical to that of OMA). 
In addition, for the performance of user 1, 
the session error probability is considered, 
which is decided by $\beta_{1,1}$ and $\beta_{2,1}$ that are 
independent of the other users' channel gains.
As a result, we do not specify any values of 
$\varphi_k$, $k = 2, \ldots, K$, as they are not needed
for the performance of user 1 in terms of the session error probability.

\subsection{Results of Symmetric NOMA}

\begin{figure}[thb]
\begin{center}
\includegraphics[width=\figwidth]{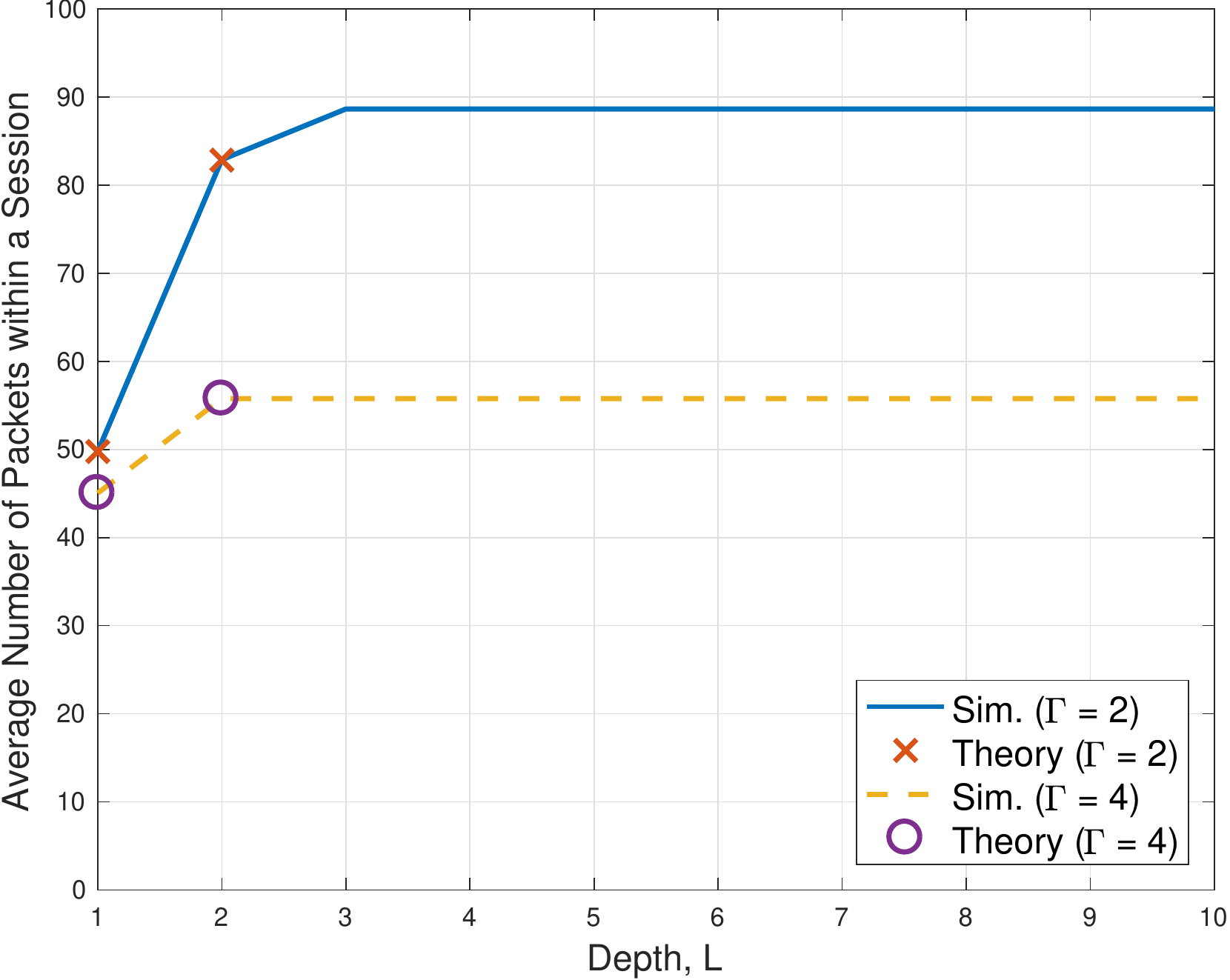} 
\end{center}
\caption{Average number of transmitted
packets in a session time (i.e., $W_{\rm S}$ slots)
for different values of depth, $L$,
when $\Gamma \in \{2, 4\}$, $\Omega = 20$, $W = 50$, and
$W_{\rm S} = 55$.}
        \label{Fig:plt0}
\end{figure}

In symmetric NOMA, the depth, $L$, becomes
the maximum number of packets that a user can transmit
in a slot (under the assumption that $L \le K = M$). 
Thus, as $L$ increases, it is expected that
the average number of transmitted packets increases
according to Lemma~\ref{L:bN}.
Fig.~\ref{Fig:plt0} shows 
the average number of transmitted
packets, $\bar N_L$,
in a session time (i.e., $W_{\rm S}$ slots)
for different values of depth, $L$,
when $\Gamma \in \{2, 4\}$, $\Omega = 20$, $W = 50$ (packets), and
$W_{\rm S} = 55$ (slots).
It is shown that although $L$ increases, 
$\bar N_L$ becomes saturated due to a high value of
$\rho_l$, $l \ge 3$. 
Thus, in most cases, $L = 2$ (i.e.,
two power levels) becomes a reasonable choice
unless the required SINR, $\Gamma$, is 
sufficient low or the power budget, $\Omega$, is extremely high,
which is however impractical.

\begin{figure}[thb]
\begin{center}
\includegraphics[width=\figwidth]{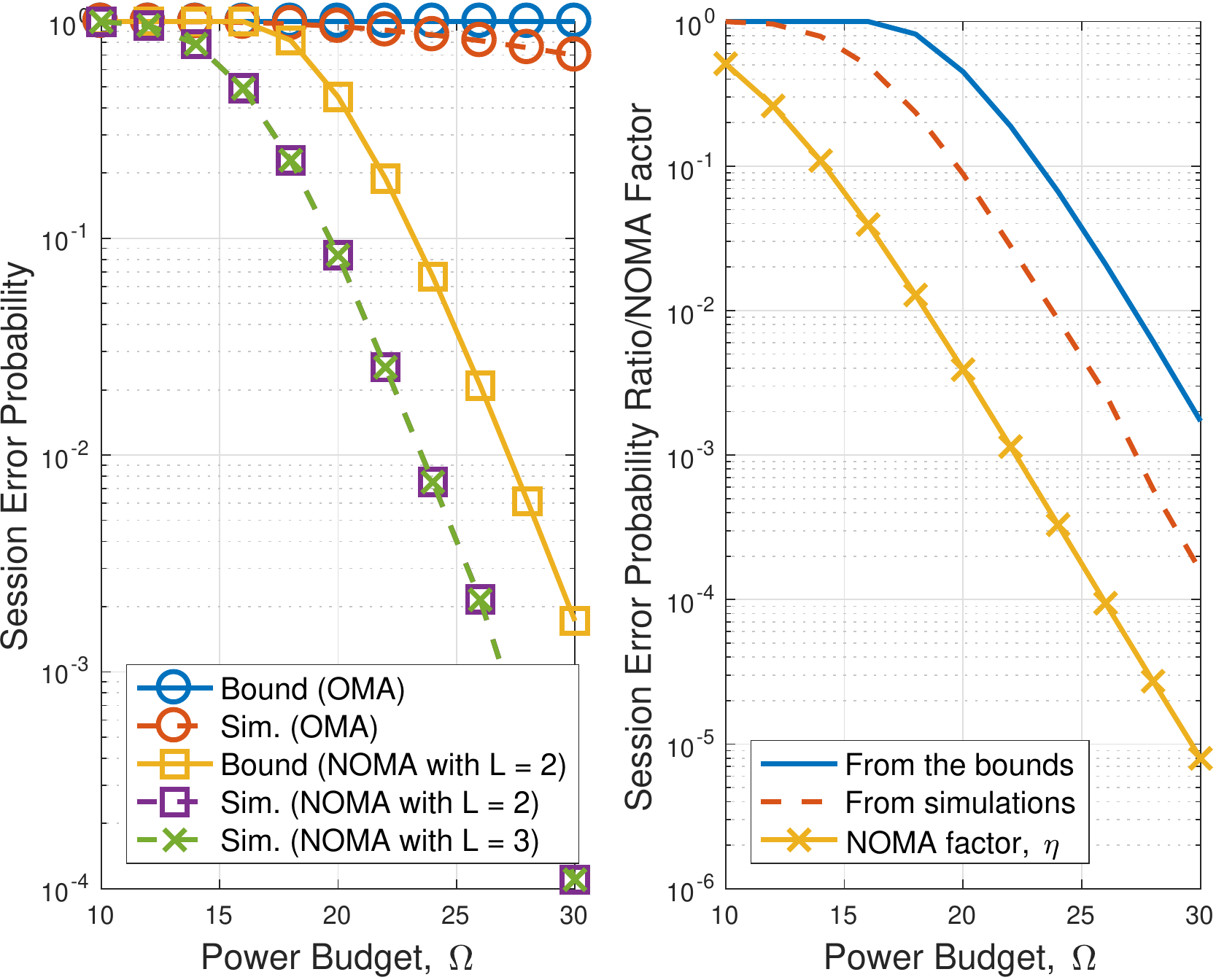} \\
\hskip 0.5cm (a) \hskip 3.5cm (b)
\end{center}
\caption{Session error probabilities and their ratio/NOMA factor 
as functions of power budget, $\Omega$,
when $\Gamma = 4$, $W = 50$, and $W_{\rm S} = 55$:
(a) session error probability versus $\Omega$;
(b) session error probability ratio/NOMA factor versus $\Omega$.}
        \label{Fig:plt1}
\end{figure}

Fig.~\ref{Fig:plt1}
shows the session error probabilities and their ratio/NOMA factor 
as functions of power budget, $\Omega$,
when $\Gamma = 4$, $W = 50$, and $W_{\rm S} = 55$.
It is shown that 
the improvement of
the session error probability of OMA is slow as $\Omega$ increases.
However, 
the session error probability significantly
decreases with $\Omega$ if opportunistic NOMA mode
is employed, which clearly
demonstrates that opportunistic NOMA mode
is an attractive scheme for SMDDC
over fading channels.
In Fig.~\ref{Fig:plt1} (a), it is shown that
the bound in \eqref{EQ:CB_L2}
can successfully predict the decreases of the
session error probability when opportunistic NOMA mode
is used. In addition, we also see that
the performance with $L = 2$ is almost the same
as that with $L = 3$, which means that the depth $L = 2$
is sufficient to take advantage of 
opportunistic NOMA mode.
In Fig.~\ref{Fig:plt1} (b), 
the NOMA factor, $\eta$, in \eqref{EQ:eta}
is shown with 
the session error probability ratio from simulation 
results
(which is represented by the dashed line).

\begin{figure}[thb]
\begin{center}
\includegraphics[width=\figwidth]{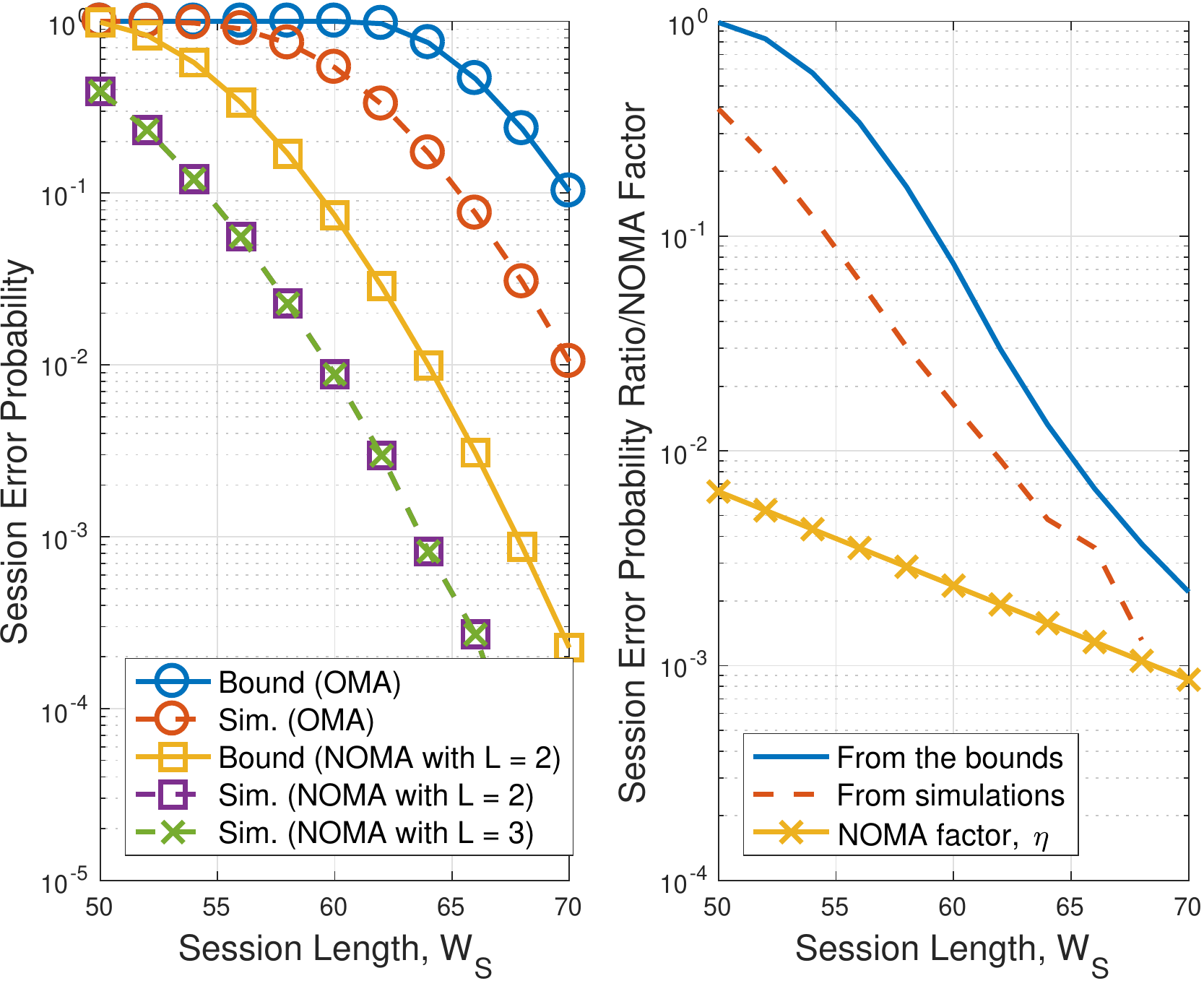} \\
\hskip 0.5cm (a) \hskip 3.5cm (b)
\end{center}
\caption{Session error probabilities and their ratio/NOMA factor 
as functions of session length, $W_{\rm S}$,
when $\Gamma = 4$, $W = 50$, and $\Omega = 20$:
(a) session error probability versus $W_{\rm S}$;
(b) session error probability ratio/NOMA factor versus $W_{\rm S}$.}
        \label{Fig:plt2}
\end{figure}

Fig.~\ref{Fig:plt2}
shows the session error probabilities and their ratio/NOMA factor 
as functions of session length, $W_{\rm S}$,
when $\Gamma = 4$, $W = 50$, and $\Omega = 20$.
It is clearly shown that the increase
of $W_{\rm S}$ decreases the session error probability
at the cost of increasing delay.

\begin{figure}[thb]
\begin{center}
\includegraphics[width=\figwidth]{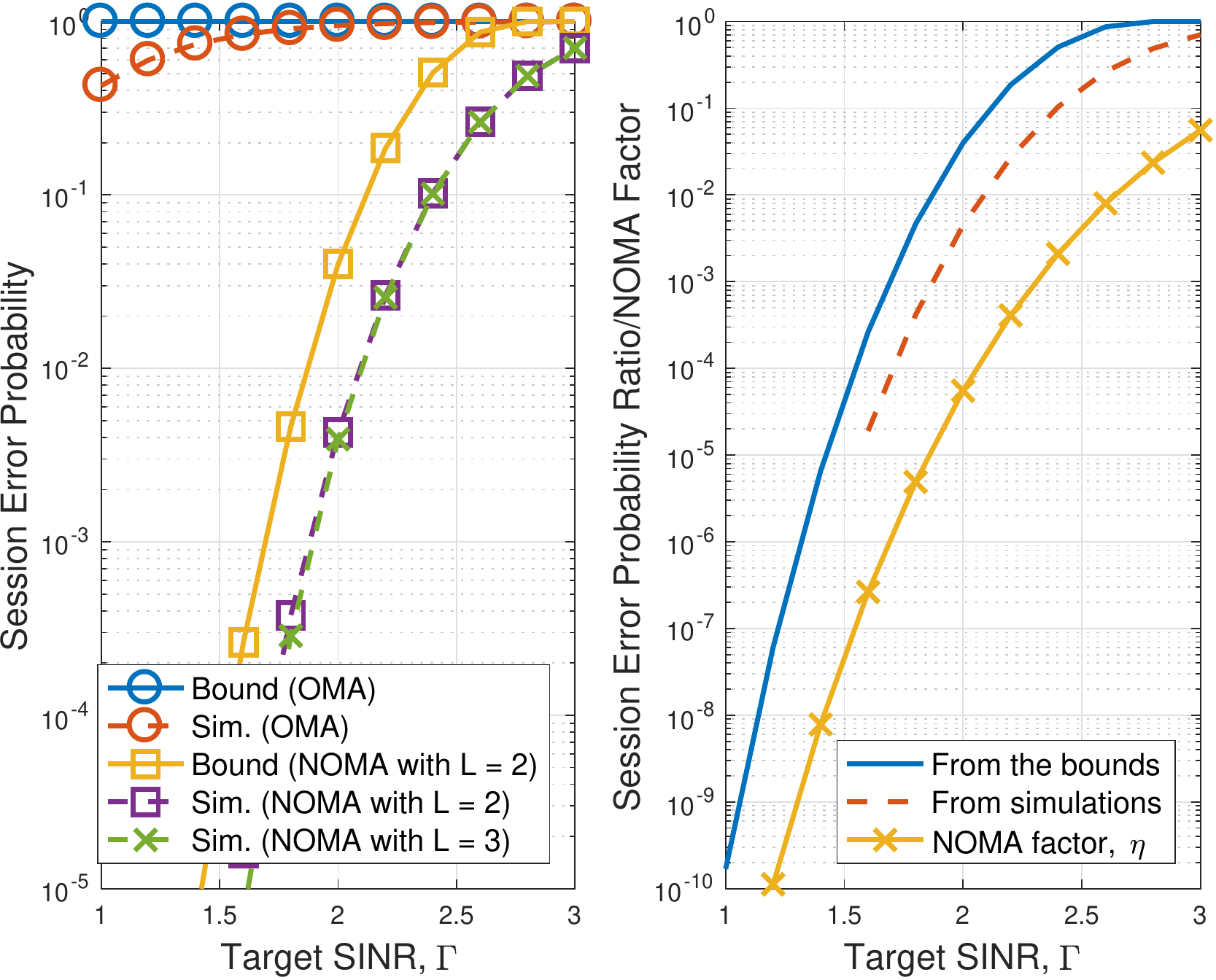} \\
\hskip 0.5cm (a) \hskip 3.5cm (b)
\end{center}
\caption{Session error probabilities and their ratio/NOMA factor 
as functions of target SINR, $\Gamma$,
when $\Omega = 10$, $W = 50$, and $W_{\rm S} = 55$:
(a) session error probability versus $\Gamma$;
(b) session error probability ratio/NOMA factor versus $\Gamma$.}
        \label{Fig:plt3}
\end{figure}

Fig.~\ref{Fig:plt3} shows 
the session error probabilities and their ratio/NOMA factor 
as functions of target SINR, $\Gamma$,
when $\Omega = 10$, $W = 50$, and $W_{\rm S} = 55$.
As the target SINR decreases, 
the session error probabilities decrease.
However, since
the code or transmission rate 
decreases with 
the target SINR,
the target SINR cannot be low.
With $\Gamma = 2$, we can see that 
the session error probability of OMA is almost 1,
while 
that with opportunistic NOMA mode 
becomes sufficiently low (i.e., less than $10^{-2}$).
This again shows that 
opportunistic NOMA mode can play a key role in SMDDC
as it can make 
the session error probability sufficiently low with a reasonably
delay constraint.

\subsection{Results of Asymmetric NOMA}

In this subsection, we present
simulation results of asymmetric NOMA
with SDO-NOMA and FO-NOMA.

\begin{figure}[thb]
\begin{center}
\includegraphics[width=\figwidth]{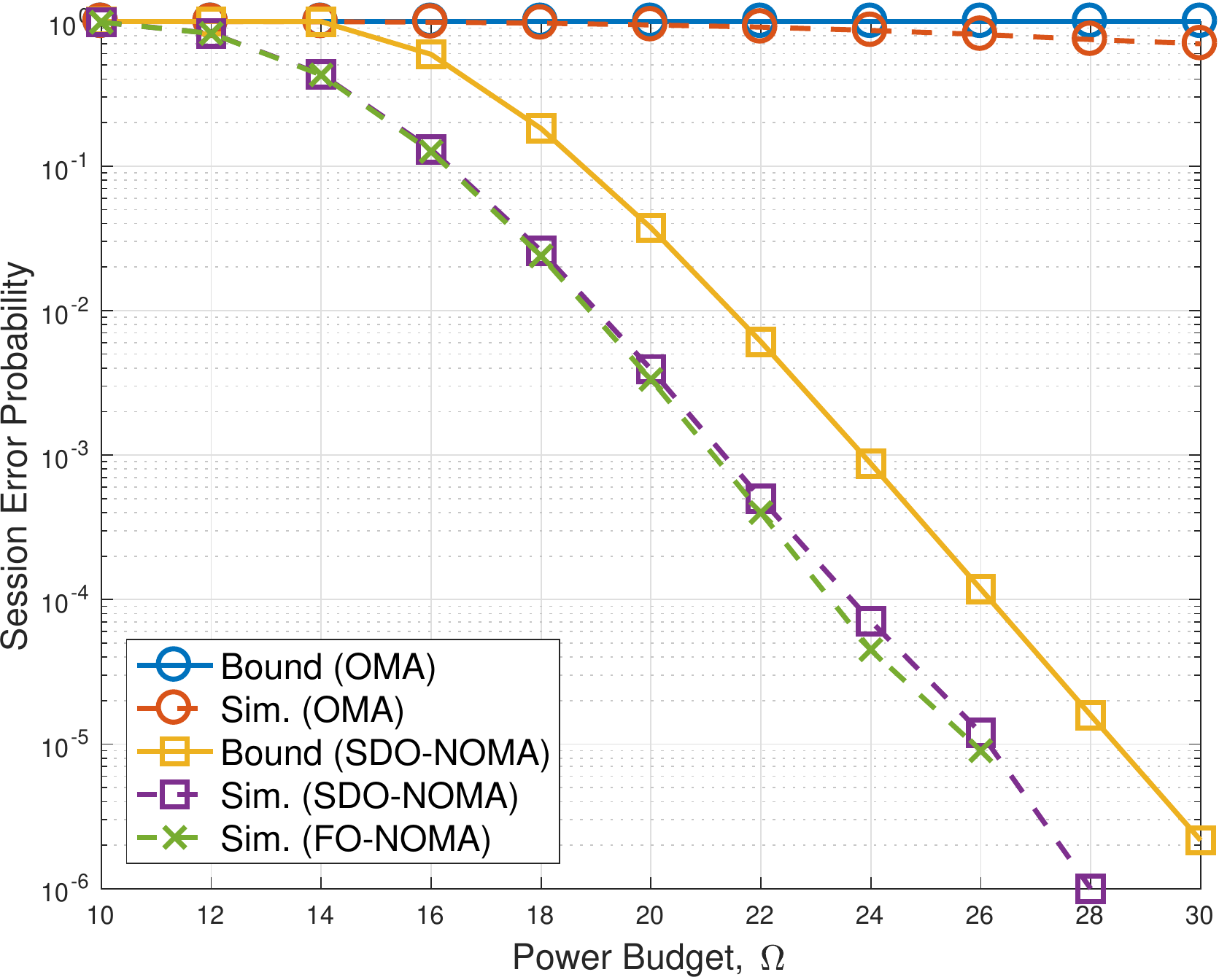} 
\end{center}
\caption{Session error probabilities of SDO-NOMA
and FO-NOMA
as functions of power budget, $\Omega$,
when $\Gamma = 4$, $K = 3$, $W = 50$, and $W_{\rm S} = 55$.}
        \label{Fig:fo_plt1}
\end{figure}

Fig.~\ref{Fig:fo_plt1}
shows the session error probabilities of SDO-NOMA
and FO-NOMA
as functions of power budget, $\Omega$,
when $\Gamma = 4$, $K = 3$, $W = 50$, and $W_{\rm S} = 55$.
From simulation results
(with the two dashed lines),
we can see that there is no significant
performance difference between 
SDO-NOMA and FO-NOMA, which means that
with a reasonable power budget,
it is unlikely to transmit more than two packets
using FO-NOMA mode.

\begin{figure}[thb]
\begin{center}
\includegraphics[width=\figwidth]{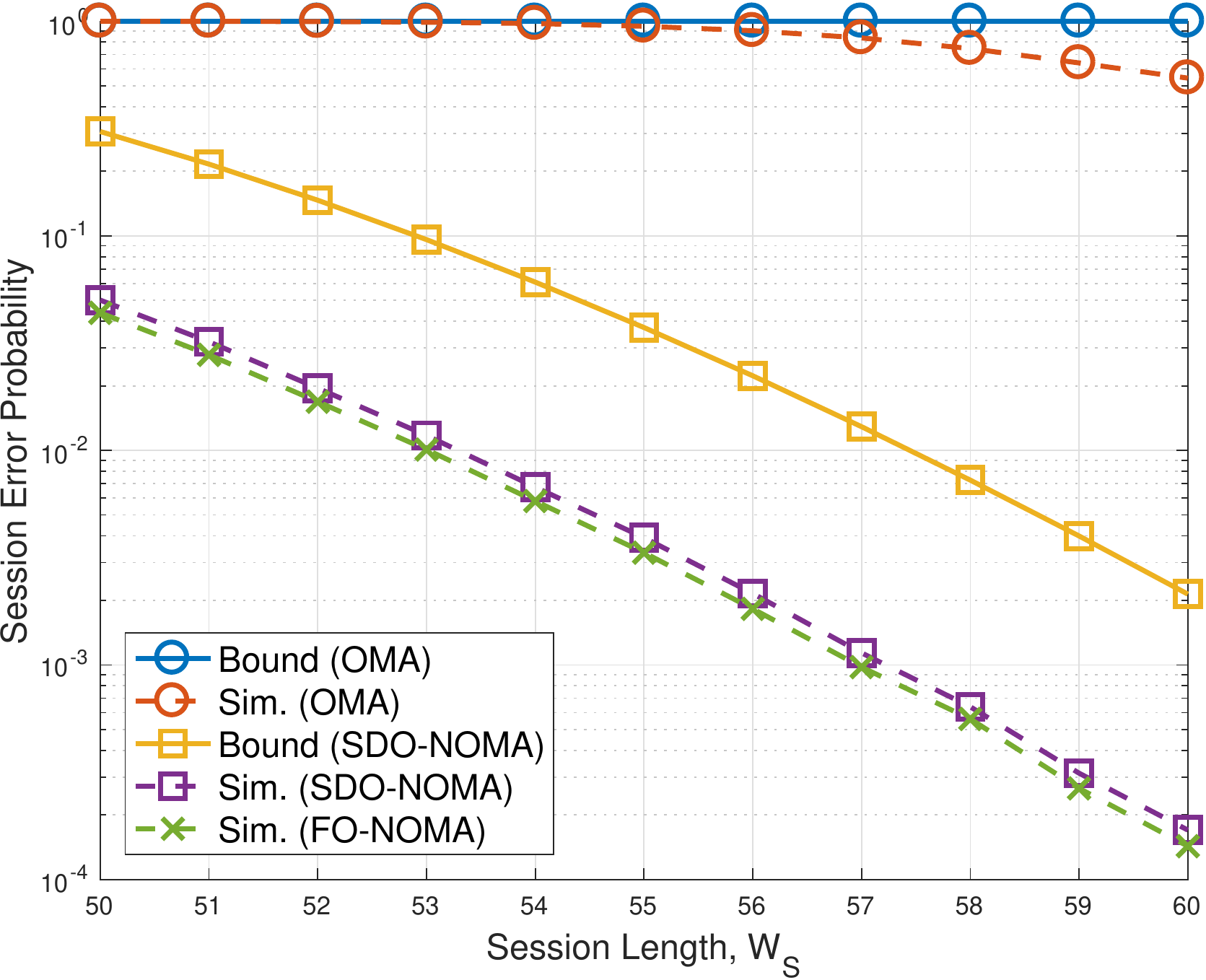} 
\end{center}
\caption{Session error probabilities of SDO-NOMA and FO-NOMA
as functions of session length, $W_{\rm S}$,
when $\Gamma = 4$, $\Omega = 15$, $K = 3$, and $W = 50$.}
        \label{Fig:fo_plt3}
\end{figure}

The session error probability versus $W_{\rm S}$
is illustrated in Fig.~\ref{Fig:fo_plt3}
when $\Gamma = 4$, $\Omega = 15$, $K = 3$, and $W = 50$.
It is noteworthy that 
even if $W_{\rm S} = W = 50$,
SDO-NOMA and FO-NOMA can provide a
low session error probability, which is about $0.05$,
while the session error probability of OMA is 1.
With $W_{\rm S} = 60$, the session error probability
of SDO-NOMA or FO-NOMA can approach $10^{-4}$.

\begin{figure}[thb]
\begin{center}
\includegraphics[width=\figwidth]{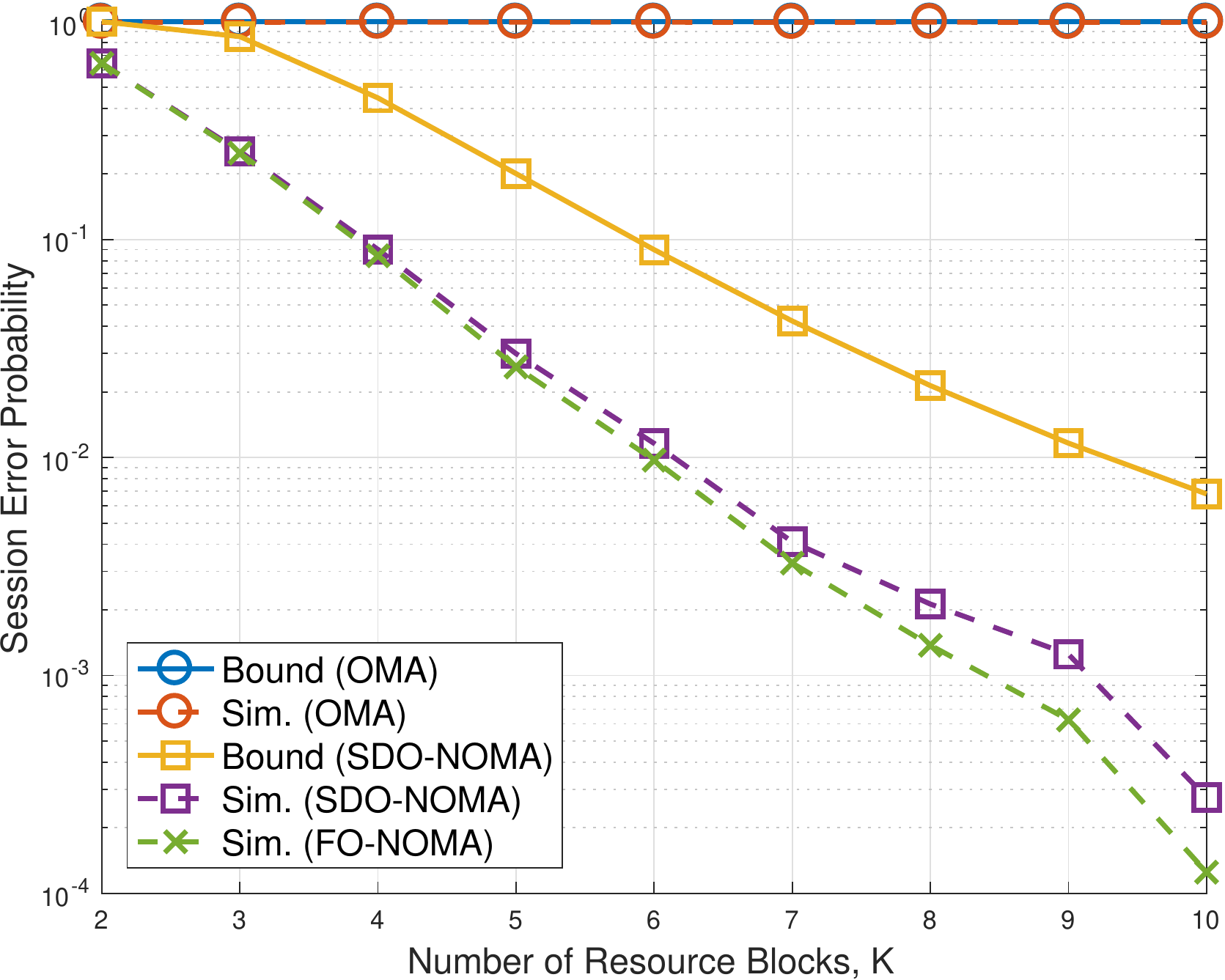} 
\end{center}
\caption{Session error probabilities of SDO-NOMA and FO-NOMA
as functions of the number of radio resource blocks, $K$,
when $\Gamma = 4$, $\Omega = 15$, $W = 50$, and $W_{\rm S} = 55$.}
        \label{Fig:fo_plt2}
\end{figure}

Fig.~\ref{Fig:fo_plt2}
shows
the session error probabilities of SDO-NOMA and FO-NOMA
as functions of the number of radio resource blocks, $K$,
when $\Gamma = 4$, $\Omega = 15$, $W = 50$, and $W_{\rm S} = 55$.
We can see
that the session error probability decreases with $K$ in
SDO-NOMA due to the increase of the selection diversity gain
and in
FO-NOMA due to the increase of radio resource blocks to 
transmit more packets per slot.
Note that although FO-NOMA can transmit more packets
than SDO-NOMA as $K$ increases (FO-NOMA can transmit up to $K$
packets per slot,
while SDO-NOMA can transmit up to two packets
per slot), there is no significant performance
difference between SDO-NOMA
and FO-NOMA in terms of the session error probability.
This implies that
the impact of 
the selection diversity gain 
in SDO-NOMA on the session error probability is 
similar to that of up to $K-1$ transmissions 
per slot.

\section{Concluding Remarks}	\label{S:Conc}

In this paper, we studied opportunistic NOMA mode
for SMDDC.
It was assumed that each user has
a set of $W$ packets that are to be transmitted
within $W_{\rm S}$ slots for SMDDC.
With OMA, it was shown that $W_{\rm S}$ has to be 
larger than $W$ for a low session error probability,
which results in a long delay.
On the other hand, it was shown that
opportunistic NOMA mode can dramatically lower the session error
probability
compared with OMA. 
In particular, under independent
Rayleigh fading, with $(W, W_{\rm S}) = (50, 60)$,
it was shown that the session error probability
of opportunistic NOMA 
can approach $10^{-4}$, while that of OMA is 0.5.
It was also confirmed by the derived NOMA factor that
shows the session error probability of opportunistic NOMA 
can be significantly lower than that of OMA
although $W_{\rm S}$ is not significantly larger than $W$. 

There are issues to be investigated in the future.
Although an upper-bound 
on the session error probability was found as a closed-form expression
to see the behavior of the session error probability,
it was noted that there is a gap between the upper-bound
and simulation results. Thus, it is desirable to find a tighter
bound in the future.
In addition, it is necessary to study the impact of imperfect
CSI on SIC, which results in degraded performance 
(in the paper, we assumed no errors in SIC 
thanks to perfect CSI).

\bibliographystyle{ieeetr}
\bibliography{noma}

\begin{thebibliography}{10}

\bibitem{Dai15}
L.~Dai, B.~Wang, Y.~Yuan, S.~Han, C.~I, and Z.~Wang, ``Non-orthogonal multiple
  access for {5G}: solutions, challenges, opportunities, and future research
  trends,'' {\em IEEE Communications Magazine}, vol.~53, pp.~74--71, September
  2015.

\bibitem{Ding_CM}
Z.~Ding, Y.~Liu, J.~Choi, M.~Elkashlan, C.~L. I, and H.~V. Poor, ``Application
  of non-orthogonal multiple access in {LTE} and {5G} networks,'' {\em IEEE
  Communications Magazine}, vol.~55, pp.~185--191, February 2017.

\bibitem{Choi17_ISWCS}
J.~Choi, ``{NOMA}: Principles and recent results,'' in {\em 2017 International
  Symposium on Wireless Communication Systems (ISWCS)}, pp.~349--354, Aug 2017.

\bibitem{Choi17_JCN}
J.~Choi, ``On generalized downlink beamforming with {NOMA},'' {\em J.
  Communications and Networks}, vol.~19, pp.~319--328, August 2017.

\bibitem{Dai18}
L.~{Dai}, B.~{Wang}, Z.~{Ding}, Z.~{Wang}, S.~{Chen}, and L.~{Hanzo}, ``A
  survey of non-orthogonal multiple access for {5G},'' {\em IEEE Communications
  Surveys Tutorials}, vol.~20, pp.~2294--2323, thirdquarter 2018.

\bibitem{Saito13}
Y.~Saito, Y.~Kishiyama, A.~Benjebbour, T.~Nakamura, A.~Li, and K.~Higuchi,
  ``Non-orthogonal multiple access ({NOMA}) for cellular future radio access,''
  in {\em Vehicular Technology Conference (VTC Spring), 2013 IEEE 77th},
  pp.~1--5, June 2013.

\bibitem{Kim13}
B.~Kim, S.~Lim, H.~Kim, S.~Suh, J.~Kwun, S.~Choi, C.~Lee, S.~Lee, and D.~Hong,
  ``Non-orthogonal multiple access in a downlink multiuser beamforming
  system,'' in {\em MILCOM 2013 - 2013 IEEE Military Communications
  Conference}, pp.~1278--1283, Nov 2013.

\bibitem{Choi08G}
J.~Choi, ``{H-ARQ} based non-orthogonal multiple access with successive
  interference cancellation,'' in {\em IEEE GLOBECOM 2008 - 2008 IEEE Global
  Telecommunications Conference}, pp.~1--5, Nov 2008.

\bibitem{Choi17}
J.~Choi, ``Effective capacity of {NOMA} and a suboptimal power control policy
  with delay {QoS},'' {\em IEEE Trans. Communications}, vol.~65,
  pp.~1849--1858, April 2017.

\bibitem{Shrouf14}
F.~{Shrouf}, J.~{Ordieres}, and G.~{Miragliotta}, ``Smart factories in industry
  4.0: A review of the concept and of energy management approached in
  production based on the internet of things paradigm,'' in {\em 2014 IEEE
  International Conference on Industrial Engineering and Engineering
  Management}, pp.~697--701, Dec 2014.

\bibitem{Dixit15}
M.~{Dixit}, J.~{Kumar}, and R.~{Kumar}, ``Internet of things and its
  challenges,'' in {\em 2015 International Conference on Green Computing and
  Internet of Things (ICGCIoT)}, pp.~810--814, Oct 2015.

\bibitem{Lom16}
M.~{Lom}, O.~{Pribyl}, and M.~{Svitek}, ``Industry 4.0 as a part of smart
  cities,'' in {\em 2016 Smart Cities Symposium Prague (SCSP)}, pp.~1--6, May
  2016.

\bibitem{Shafiq12}
M.~Z. Shafiq, L.~Ji, A.~X. Liu, J.~Pang, and J.~Wang, ``A first look at
  cellular machine-to-machine traffic: Large scale measurement and
  characterization,'' {\em SIGMETRICS Perform. Eval. Rev.}, vol.~40,
  pp.~65--76, June 2012.

\bibitem{Sehati18}
A.~{Sehati} and M.~{Ghaderi}, ``Online energy management in iot applications,''
  in {\em IEEE INFOCOM 2018 - IEEE Conference on Computer Communications},
  pp.~1286--1294, April 2018.

\bibitem{Naik17}
N.~{Naik}, ``Choice of effective messaging protocols for {IoT} systems: {MQTT},
  {CoAP}, {AMQP} and {HTTP},'' in {\em 2017 IEEE International Systems
  Engineering Symposium (ISSE)}, pp.~1--7, Oct 2017.

\bibitem{Kim18}
D.~{Kim}, H.~{Lee}, and D.~{Kim}, ``Enhanced industrial message protocol for
  real-time {IoT} platform,'' in {\em 2018 International Conference on
  Electronics, Information, and Communication (ICEIC)}, pp.~1--2, Jan 2018.

\bibitem{TseBook05}
D.~Tse and P.~Viswanath, {\em Fundamentals of Wireless Communication}.
\newblock Cambridge University Press, 2005.

\bibitem{LinBook}
S.~Lin and D.~J. Costello, Jr, {\em Error Control Coding: Fundamentals and
  Applications}.
\newblock Englewood Cliffs, N.J.: Prentice Hall, 1983.

\bibitem{Liu18}
Y.~{Liu}, M.~{Derakhshani}, and S.~{Lambotharan}, ``Outage analysis and power
  allocation in uplink non-orthogonal multiple access systems,'' {\em IEEE
  Communications Letters}, vol.~22, pp.~336--339, Feb 2018.

\bibitem{Cui18}
J.~{Cui}, Z.~{Ding}, and P.~{Fan}, ``Outage probability constrained
  {MIMO}-{NOMA} designs under imperfect {CSI},'' {\em IEEE Trans. Wireless
  Communications}, vol.~17, pp.~8239--8255, Dec 2018.

\bibitem{Choi14}
J.~Choi, ``Non-orthogonal multiple access in downlink coordinated two-point
  systems,'' {\em IEEE Commun. Letters}, vol.~18, pp.~313--316, Feb. 2014.

\bibitem{Ding14}
Z.~Ding, Z.~Yang, P.~Fan, and H.~Poor, ``On the performance of non-orthogonal
  multiple access in {5G} systems with randomly deployed users,'' {\em IEEE
  Signal Process. Letters}, vol.~21, pp.~1501--1505, Dec 2014.

\bibitem{CoverBook}
T.~M. Cover and J.~A. Thomas, {\em Elements of Information Theory}.
\newblock NJ: John Wiley, second~ed., 2006.

\bibitem{MacKayBook}
D.~J.~C. MacKay, {\em Information Theory, Inference, and Learning Algorithms}.
\newblock Cambridge University Press, 2003.

\bibitem{Polyanskiy10IT}
Y.~Polyanskiy, H.~V. Poor, and S.~Verdu, ``Channel coding rate in the finite
  blocklength regime,'' {\em IEEE Trans. Information Theory}, vol.~56,
  pp.~2307--2359, May 2010.

\bibitem{Durisi16}
G.~Durisi, T.~Koch, and P.~Popovski, ``Toward massive, ultrareliable, and
  low-latency wireless communication with short packets,'' {\em Proceedings of
  the IEEE}, vol.~104, pp.~1711--1726, Sept 2016.

\bibitem{DavidBook}
H.~A. David, {\em Order Statistics}.
\newblock New York: John Wiley \& Sons, 1980.

\bibitem{Mitz05}
M.~Mitzenmacher and E.~Upfal, {\em Probability and Computing: Randomized
  Algorithms and Probability Analysis}.
\newblock Cambridge University Press, 2005.

\bibitem{Arratia89}
R.~Arratia and L.~Gordon, ``Tutorial on large deviations for the binomial
  distribution,'' {\em Bulletin of Mathematical Biology}, vol.~51, no.~1,
  pp.~125 -- 131, 1989.

\bibitem{Dembo98}
A.~Dembo and O.~Zeitouni, {\em Large Deviations Techniques and Applications}.
\newblock Applications of mathematics, Springer, 1998.

\bibitem{Bullen13}
P.~Bullen, {\em Handbook of Means and Their Inequalities}.
\newblock Mathematics and Its Applications, Springer Netherlands, 2013.

\bibitem{Schaible81}
S.~Schaible, ``Fractional programming: Applications and algorithms,'' {\em
  European Journal of Operational Research}, vol.~7, no.~2, pp.~111 -- 120,
  1981.
\newblock Fourth EURO III Special Issue.

\end{thebibliography}

\end{document}